\documentclass[arxiv,10pt,usenames,dvipsnames]{jmlr} %

\usepackage{lipsum}
\usepackage{mathtools}

\usepackage[backend=biber,style=numeric,sortcites,sorting=none,backref,natbib,hyperref]{biblatex}
\usepackage{xcolor} %

\usepackage{tikz}
\usetikzlibrary{positioning}

\usepackage{multirow}

\usepackage{colortbl}
\usepackage[normalem]{ulem}
\useunder{\uline}{\ul}{}

\usepackage{booktabs}
\usepackage{siunitx}

\DeclareMathOperator{\EX}{\mathbb{E}}%

\usepackage[acronym,nowarn]{glossaries}
\makeglossaries
\newacronym{opm}{OPM}{outcome prediction model}
\glsdisablehyper

\theorembodyfont{\upshape}
\theoremheaderfont{\scshape}
\theorempostheader{:}
\theoremsep{\newline}

\jmlrvolume{}
\jmlryear{2024}
\jmlrsubmitted{}
\jmlrpublished{}

\title[harmful self-fulfilling prophecies]{When accurate prediction models yield harmful self-fulfilling prophecies}

\author{%
\Name{Wouter A.C. {van Amsterdam, MD, PhD}\nametag{\thanks{these authors contributed equally}}} \Email{w.a.c.vanamsterdam-3@umcutrecht.nl}\\
\addr Department of Data Science and Biostatistics \\
Julius Center of Health Sciences and Primary Care \\
University Medical Center Utrecht, Utrecht, the Netherlands\\
University of Utrecht, Utrecht, the Netherlands\\
Heidelberglaan 100, 3584 CX Utrecht, the Netherlands \\
corresponding author
\AND
\Name{Nan {van Geloven, PhD}} \\
\addr Department of Biomedical Data Sciences\\
Leiden University Medical Center, Leiden, the Netherlands
\AND
\Name{Jesse H. Krijthe, PhD} \\
\addr Pattern Recognition \& Bioinformatics\\
Delft University of Technology, Delft, the Netherlands
\AND
\Name{Rajesh Ranganath, PhD} \\
\addr 
Courant Institute of Mathematical Science, Department of Computer Science \\    
Center for Data Science \\
New York University, New York City, USA
\AND
\Name{Giovanni Cin\`a$^*$, PhD} \Email{g.cina@amsterdamumc.nl}\\
\addr Department of Medical Informatics\\
Amsterdam University Medical Center, Amsterdam, the Netherlands\\
Institute for Logic, Language and Computation\\
University of Amsterdam, Amsterdam, the Netherlands\\
Pacmed, Amsterdam, the Netherlands
}

\begin{document}

\maketitle

\pagebreak

\begin{abstract}
Prediction models are popular in medical research and practice. By predicting an outcome of interest for specific patients, these models may help inform difficult treatment decisions, and are often hailed as the poster children for personalized, data-driven healthcare.
We show however, that using prediction models for decision making can lead to harmful decisions, even when the predictions exhibit good discrimination after deployment. These models are harmful self-fulfilling prophecies: their deployment harms a group of patients but the worse outcome of these patients does not invalidate the predictive power of the model. Our main result is a formal characterization of a set of such prediction models. Next we show that models that are well calibrated before and after deployment are useless for decision making as they made no change in the data distribution. These results point to the need to revise standard practices for validation, deployment and evaluation of prediction models that are used in medical decisions.
 
\end{abstract}

\section{Introduction}
\label{sec:intro}

Clinicians and medical researchers frequently employ \glspl{opm}: statistical models that predict a certain medical outcome based on a patient's characteristics \citep{steyerberg2009applications}.
Researchers develop \glspl{opm} to provide information to clinicians so they may use this information in difficult treatment decisions (e.g. \citet{salazar_gene_2011}).
In some cases, clinicians will treat patients with a bad expected outcome more aggressively, for example by giving cholesterol lowering medication to patients with a high predicted risk of a heart attack \citep{arnett_2019_2019,karmali_blood_2018}.
In other cases, for instance when the treatment is burdensome or scarcely available (e.g. ventilator machines on the intensive care during a pandemic), clinicians may reserve treatment for patients with a good predicted outcome.

Many such \glspl{opm} are added to the protocol of care by designing specific thresholds for specific actions \citep{arnett_2019_2019}.
If the predicted outcome is above or below the threshold a certain action is taken, e.g. the patient receives a more aggressive therapy.
The basis for including an \gls{opm} in a care protocol is generally predictive accuracy in validation studies \citep{kattan_american_2016}.
In these validation studies, the \gls{opm} may or may not have been used to inform treatment decisions.
While the difference between a clinical trial of an OPM's deployment and evaluation of performance metrics is appreciated in the medical literature, there are still notable examples where the latter is perceived to be sufficient to justify the implementation of OPMs in the protocol of care. This is reflected in several guidelines and reviews \citep{kattan_american_2016, rahimi2014risk}.

At first, it may seem that using OPMs for decision support is beneficial since giving more information should lead to better treatment decisions.
However, implementing a prediction model for treatment decisions is an intervention that changes treatment decisions and thus patient outcomes.
Whether this change in treatment policy improves patient outcomes is not determined by prediction accuracy in a validation study \citep{vanamsterdamAlgorithmsActionImproving2024}.
For instance, in cases where a certain patient subpopulation historically received suboptimal care, an accurate \gls{opm} will predict a worse outcome for these patients compared to similar patients of a different subpopulation.
If clinicians decide to withhold effective treatments (e.g., due to scarcity or perceived futility) to this underserved subpopulation based on the \gls{opm}'s prediction of a bad outcome, the implementation of the \gls{opm} perpetuated biases or caused harm to these patients, despite its accuracy.
Moreover, the implementation of this harmful new policy brought about the scenario predicted by the \gls{opm}, as in a \emph{self-fulfilling prophecy}.
One concrete example where clinicians treat patients with a bad expected outcome less aggressively is in small cell lung cancer.
Prognostic scores for small cell lung cancer patients, such as the Manchester score \cite{cernyPretreatmentPrognosticFactors1987} are specifically intended to not over-treat patients with a bad predicted outcome because this is expected to be futile \cite{hagmannValidationPretreatmentPrognostic2022,ferraldeschiModernManagementSmallCell2007}.

Recognizing that prediction model performance may change over time, across health care settings and in certain patient subgroups, many call for an increased monitoring of AI models with model updating mentioned as the best approach \cite{shahNationwideNetworkHealth2024,celiSourcesBiasArtificial2022,futomaMythGeneralisabilityClinical2020}.
However, these approaches fall short as they put the wrong metric upfront: prediction accuracy. We show that the value of a prediction model is not directly derived from its accuracy and in some cases having worse prediction accuracy over time is exactly what we want from a patient outcomes perspective. Focusing on predictive performance only might lead to the employment of a new policy that is harmful for patients, or to the unduly withdrawal of a policy that was in fact beneficial. 

In this article we address the following questions: 1) Under what conditions is a new policy based on an \gls{opm} going to be harmful, meaning that it leads to worse outcomes than before using the model?
2) In what circumstances would such a harmful policy go undetected by measures of discrimination or calibration? 
In what follows we provide a formalization of the case where patients with a high predicted probability of the outcome get treatment, where the outcome may be be preferable (e.g. 1-year survival) or undesirable (e.g. a heart attack).
Specifically, we examine the setting where a new OPM is supposed to `personalize' an  existing treatment policy by considering additional features. 
Section \ref{sec:notation} provides a motivating example, notation and definitions,
Section \ref{sec:theorems} presents the main results concerning \glspl{opm} that are harmful and self-fulfilling prophecies.
We first show that even in a simple setup with a binary covariate, a non-trivial subset of \glspl{opm} yields harmful self-fulfilling prophecies. This means that such models cause harm but exhibit good discrimination on post-deployment data, meaning that naively interpreting this as a successful deployment leads to harmful policies. These theoretical results are paired with numerical experiments demonstrating that harmful self-fulfilling prophecies can occur without assuming extreme treatment effects or treatment effect heterogeneity.
Next, perhaps surprisingly, we show that when an \gls{opm} is well calibrated on both 1) the historical data and 2) a validation study where the model is used for treatment decisions, the \gls{opm} is not useful for decision making.

Based on our results, several common practices in building and deploying \glspl{opm} intended for decision making need revision:
1. Developing \glspl{opm} on observational data without regard of the historical treatment policy is potentially dangerous, because the change in treatment policy between pre- and post-deployment is what determines the effect of the model on patient outcomes.
2. Implementing a personalized outcome prediction model is not always beneficial, even if the model is very accurate.
3. When monitoring discrimination prospectively after deployment, sometimes good discrimination means a harmful model and sometimes a beneficial one.

\section{Notation and definitions}
\label{sec:notation}

\subsection{Motivating example of a harmful self-fulfilling prophecy}
\label{example}

\begin{figure}
    \centering
    \includegraphics[width=\textwidth]{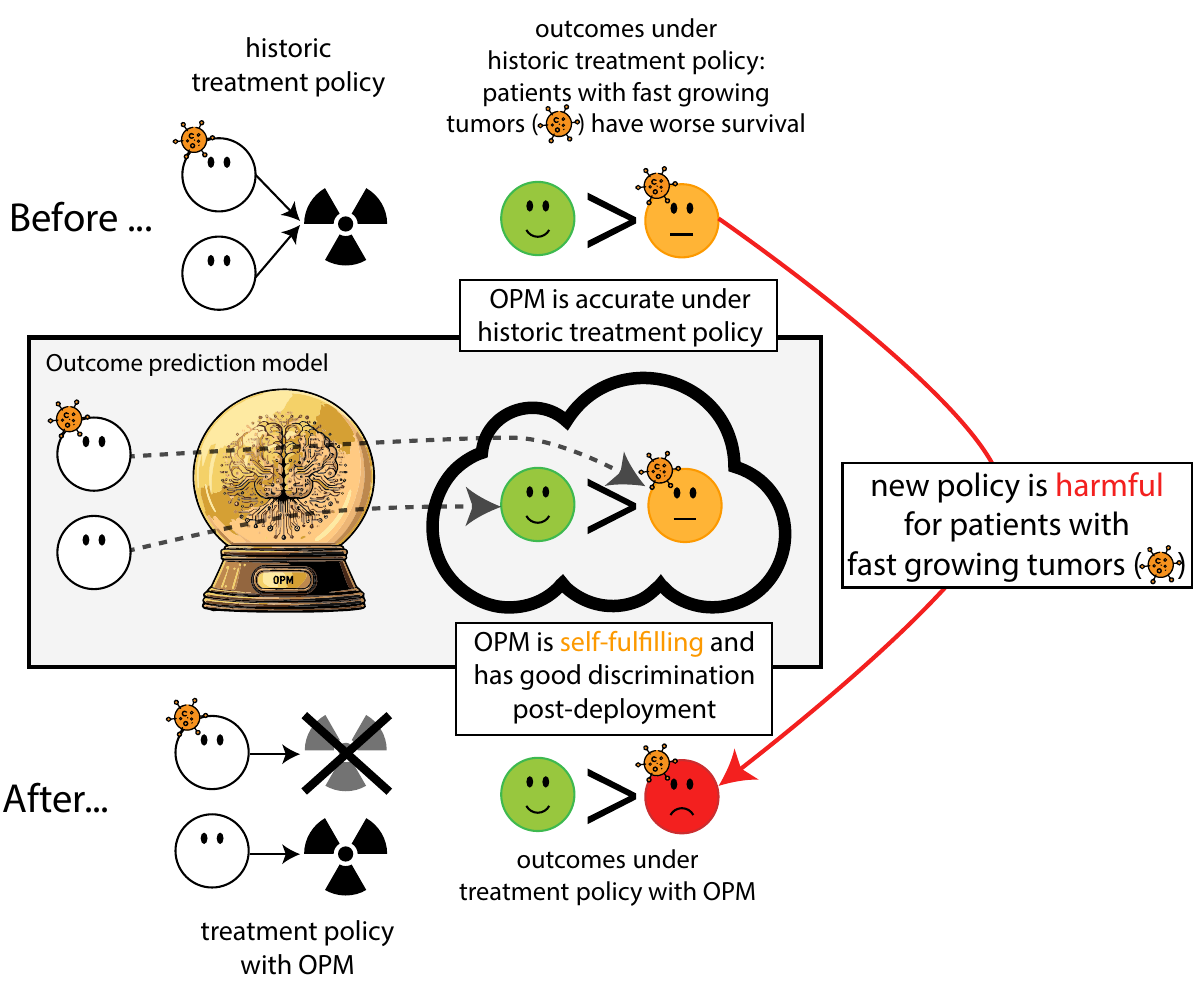}
    \caption{Some outcome prediction models yield harmful self-fulfilling prophecies when used to guide treatment decisions, meaning the new policy harms a subgroup of patients but the prediction model has good discrimination post-deployment because the patients who are harmed were already expected to have worse outcomes.}
    \label{fig:rt_example}
\end{figure}

We start with a hypothetical example based on realistic medical assumptions that would result in an OPM yielding a policy that is both harmful, meaning patient outcomes are worse compared to before deployment, and self-fulfilling, meaning the OPM has good discrimination post deployment. In Appendix \ref{app:example} we provide a formal version of this example with corresponding equations and proof, Figure \ref{fig:rt_example} provides an illustration.

Consider the problem of selecting a subset of end-stage cancer patients for palliative radiotherapy.
Such treatment has side-effects and thus domain experts advise to reduce over-treatment in this population.
To comply with this advice, a medical center needs to decide which patients will not be eligible anymore for the therapy.
The medical center decides to give the therapy to patients with the longest expected overall survival, under the assumption that for these patients the side-effects are justifiable.
To support this policy, researchers build an \gls{opm} to predict the probability of 6-months overall survival based on pre-treatment tumor growth rate using historical patient records from the medical center.
Fast-growing tumors are more aggressive so these patients have a shorter survival overall.
The medical center decides to use this model to allocate the therapy and tests the model's discrimination post deployment.

The new treatment policy with the OPM  is thus ``treat patients with slow growing tumors but not those with fast growing tumors''.
However, fast-growing tumors respond better to radiotherapy than slow growing tumors \citep{breurGrowthRateRadiosensitivity1966}. 
So the new \gls{opm}-based policy treats exactly the wrong patients: those who do not benefit from treatment still receive it, those who would benefit from treatment do not, so deployment of the model was harmful.
The contrast in survival between patients with fast-growing tumors and slow-growing tumors has gotten only more pronounced post-deployment, meaning that, paradoxically, the OPM has good discrimination before and after deployment.

This potential for deploying harmful self-fulfilling prophecies by only relying on measures of predictive discrimination is clearly undesirable. We now provide a formal framework describing when these situations occur so we can also understand how they may be prevented, revealing also a dual case where OPMs that provide benefit to patient subgroups show worse post-deployment discrimination.

\subsection{Notation and definitions}

We assume a binary treatment $T$, a binary outcome $Y$ and a binary feature $X \in \mathcal{X} = \{0,1\}$. We denote the outcome obtained with setting treatment $T$ to $t$ as  $Y_t$.
An \gls{opm} is a function trained on historical data to predict the probability of the outcome of interest. %
We use $\pi_i(X)$ to denote a policy for assigning treatment, possibly conditional on $X$, with an index $i$ to indicate what policy we are referring to. Throughout the paper $\pi_0$ will be used to indicate the \emph{historic treatment policy} that was in place in the data in which the \gls{opm} was developed.

We assume the historical policy is constant and deterministic, meaning that it is always equal to 0 or 1 (i.e. patients were always treated or never treated).
Next we define what it means to craft a policy based on an existing OPM.
We will be concerned only with \textit{threshold-based policies}, namely policies that assign treatment based on a threshold $\lambda\in \mathbb{R}$.
In our setup, policies assign treatment to patients only if the expected outcome is above $\lambda$, which could mean either a desirable outcome (e.g. 1-year survival) or undesirable (e.g. a heart attack).

\begin{definition}[Policy informed by OPM] \label{def:opm_policy}

Let $f: X \to [0,1]$ be an OPM and $\lambda\in \mathbb{R}$ a threshold. We call $\pi_f$ a \emph{policy informed by $f$} and define it as follows
\begin{equation}
	\label{eq:pif}
	\pi_f(x) = 
	\begin{cases}
		1 \hspace{1cm} f(x) > \lambda \\
		0 \hspace{1cm} f(x) \leq \lambda
	\end{cases}
\end{equation}
\end{definition}

Such policies describe the post-deployment scenario, when the OPM influences treatment assignment.
This deployment will change some of the (conditional) probability distributions compared to pre-deployment.
We distinguish probabilities pre- and post-implementation using subscripts: $p_i(.)$ with $i\in \{0,f\}$ respectively.
We now present the first key idea of this paper, namely the special class of OPMs whose predictions are realized upon implementation. We consider as a metric of discrimination the popular `Area under the ROC-curve' \cite{hanleyMeaningUseArea1982} (AUC).

\begin{definition}[Self-fulfilling OPM]\label{def:selffulfilling_auc}
Let $f: \mathcal{X} \to [0,1]$ be an OPM, $\lambda\in \mathbb{R}$ a threshold and let $\pi_f$ be the policy informed by $f$.
Let $AUC(\pi_{i})$ denote the AUC of this OPM on data generated with the historic policy ($\pi_{0}$) or with the policy defined by $f$ ($\pi_{f}$).
We call the pair $(f,\lambda)$ \emph{self-fulfilling} if the AUC remains the same or increases post-deployment, namely iff:
\begin{equation}
    \label{eq:selffulfilling_auc}
    \text{AUC}(\pi_f) \geq \text{AUC}(\pi_0)
\end{equation}
\end{definition}

Finally, we specify what we mean with an OPM being harmful in comparison with the status quo.

\begin{definition}[Harmful OPM]
	\label{def:pointwiseharmful}
    Let $f: \mathcal{X} \to [0,1]$ be an OPM, $\lambda\in \mathbb{R}$ a threshold, let $\pi_0$ denote the historic treatment policy and let $\pi_f$ be the policy informed by $f$.

    We write the expected outcomes under the different policies as\begin{equation}\label{eq:calibration_train}
        p_i(Y = 1|X) = \EX_{T \sim \pi_i(X)} p(Y_T=1|X)
    \end{equation}
    where $i=0$ denotes the historical distribution and $i=f$ the distribution under $\pi_f$.
    We call $f$ \emph{harmful} for the group with $X=x$ with $p(X=x) > 0$ if the expected outcome of the group\footnote{Note that this is different from a model being marginally harmful, i.e. applying $\pi_f$ leads to worse outcomes on average. However, we will later see that in our setup with binary $X$, one of the two groups has the same outcomes pre- and post-deployment so an OPM that is harmful to a subgroup will also be marginally harmful.} is worse under the new policy compared to the old policy, namely when $Y=1$ is preferable iff
    \begin{equation}
    \label{eq:harmful}
    p_f(Y=1|X=x) < p_0(Y=1|X=x)
\end{equation}
    or when $Y=0$ is preferable iff
    \begin{equation}
    \label{eq:harmful}
    p_f(Y=1|X=x) > p_0(Y=1|X=x)
\end{equation}
\end{definition}

When a policy informed by an OPM is both harmful and self-fulfilling we have a worst-case scenario where the new policy is causing harm to a subgroup but this, perhaps counter-intuitively, does not result in a decrease in AUC post-deployment.

\section{Results}
\label{sec:theorems}

We now move to the main results, whose proofs can be found in Appendix \ref{apd:proofs}.
The setting where a new OPM is supposed to `personalize' an already existing treatment policy by considering more features is encoded as follows: the new OPM considers a feature $X$ that was previously ignored by the historical policy, specifically $\pi_0$ is constant and deterministic. %
In addition, the new policy $\pi_f$ is not constant but varies with $X$.

\subsection{Harmful models may have good discrimination post-deployment}
We state our main observation as an informal theorem.

\begin{theorem}[Informal main result]\label{thm:main}
Let $\pi_f$ be the policy informed by the OPM $f$ using a threshold $\lambda$. Assume that: i) the historical policy $\pi_0$ is constant and deterministic ii)
the new policy $\pi_f$ is  not constant, i.e. not always equal to 1 or 0 and iii) the marginal distribution of $X$ is the same pre and post deployment: $p_i(X) = p(X)$ for $i\in \{0, f\}$.

Under these assumptions, a non-trivial subset of \glspl{opm} will demonstrate good post-deployment discrimination because they yield self-fulfilling prophecies, and at the same time their deployment harmed patients.

\end{theorem}

We proceed to characterize the contours of the subset of self-fulfilling and harmful \glspl{opm}.

\begin{proposition}[Self-fulfilling]\label{thm:selffulfilling}
Suppose the assumptions of Theorem \ref{thm:main} hold. 
Furthermore assume that the joint probabilities of $X$ and $Y$ are non-deterministic both pre- and post-deployment:
\begin{align}\label{eq:ass_positivity}
	0 < p_i(Y=1,X=x) < 1, \forall x \in \mathcal{X}
\end{align}

Then the following two statements are true: i) if the treatment effect is always positive, namely $ \forall x \in \mathcal{X}: p(Y_1=1|X=x) \geq p(Y_0=1|X=x)$, then $(f,\lambda)$ is self-fulfilling; ii) if the treatment effect is always negative, meaning $\forall x \in \mathcal{X}: p(Y_1=1|X=x) < p(Y_0=1|X=x)$, then $(f,\lambda)$ is not self-fulfilling.
\end{proposition}

\begin{remark}
    Proposition \ref{thm:selffulfilling} gives sufficient conditions for an OPM yielding a self-fulfilling prophecy.
    When $Y=1$ is preferable, meaning the new policy treats only those with a favorable predicted outcome (e.g. under resource scarcity), the sufficient condition is that the treatment effect is beneficial for all values of $X$.
    When instead $Y=0$ is preferable, meaning the `treat high-risk patients'-setting, the sufficient condition is that treatment is detrimental for all values of $X$. 
    Treatments that are always detrimental are less likely to be used in practice as most often treatments are approved for use after they are proven to be beneficial on average with an RCT.
    In this case of `treat high risk', self-fulfilling prophecies may still occur when the treatment is detrimental to a subgroup of patients.
\end{remark}

\begin{remark}
    Proposition \ref{thm:selffulfilling} does not depend on the \gls{opm}'s discrimination in the historical data, meaning that models with `good' discrimination (i.e. high AUC) and `bad' discrimination (low AUC) are equally susceptible to yielding self-fulfilling prophecies under the conditions of the proposition.
\end{remark}

Now we know when \glspl{opm} are self-fulfilling and thus have good post-deployment discrimination, but can these self-fulfilling \glspl{opm} also be harmful?
Proposition \ref{thm:self_harmful} indicates that they can:

\begin{proposition}[Harmful] \label{thm:self_harmful}

Under the assumptions of Theorem \ref{thm:main}, when $Y=1$ is preferable, $f$ is harmful for the group with $X=x$ iff
\begin{enumerate}
    \item $\pi_0(x) = 1$ and $\pi_f(x) = 0$ and $p(Y_1=1|X=x) > p(Y_0=1|X=x)$ or
    \item $\pi_0(x) = 0$ and $\pi_f(x) = 1$ and $p(Y_1=1|X=x) < p(Y_0=1|X=x)$
\end{enumerate}
When $Y=0$ is preferable, the inequality signs reverse.
\end{proposition}

The conditions of this Proposition indicate that, as one would expect, removing the treatment from this group is harmful iff $p(Y_1=1|X=x) > p(Y_0=1|X=x)$ (assuming $Y=1$ is preferable), i.e. if the effect of the treatment was positive for this group.
Conversely, adding treatment to group with $X=x$ is damaging iff  $p(Y_1=1|X=x) < p(Y_0=1|X=x)$ (when $Y=1$ is preferable), meaning that the treatment decreases the outcome for the group.

\begin{remark}[harmful OPMs are marginally harmful]
    Under the assumptions of Theorem \ref{thm:main}, OPMs that are harmful for one subgroup are also harmful on average, as the other subgroup's treatment policy and outcomes do not change.
\end{remark}

 Taking together Proposition \ref{thm:selffulfilling} on when \glspl{opm} yield self-fulfilling prophecies and Proposition \ref{thm:self_harmful} on when \gls{opm} deployment is harmful, we reach the perhaps surprising conclusion of Theorem \ref{thm:main}: even in the simple setup of binary treatment and binary $X$, some \glspl{opm} are both self-fulfilling prophecies, and thus demonstrate good post-deployment discrimination, and harm a patient subgroup when deployed. We present the above example based on realistic medical assumptions in a formalized way in Appendix \ref{app:example}.
In Table \ref{tab:results} we list the cases in which OPM deployment is harmful, based on three pieces of information that are available post-deployment: i) is $Y=1$ preferable or undesirable? ii) was the historical policy `treat everyone' or `treat no one'? and iii) did the AUC of the OPM increase post-deployment compared to the AUC pre-deployment (i.e. is the OPM self-fulfilling)? Finally, we note that the performance of the OPM on the historical data does not feature in the assumptions or statement of Proposition \ref{thm:self_harmful}. This entails, contrary to common expectation, that a high performance on historical data, including external validation, provides no guarantee on whether the OPM-driven policy will be beneficial.

\begin{table}[]
\centering
\begin{tabular}{llll}
interpretation of $Y=1$ (and policy) & $\pi_0$ & $AUC(\pi_f) - AUC(\pi_0)$                                          & OPM deployment was \\ \hline
                                     & 0 (treat no one)       & \textgreater 0 (self-fulfilling) & \cellcolor[HTML]{FE996B}harmful \\
                                     & 0 (treat no one)       & \textless 0 (not self-fulfilling) & \cellcolor[HTML]{32CB00}beneficial \\
\multirow{-2}{*}{undesirable (treat high risk patients)} & 1 (treat everyone)       & \textgreater 0 (self-fulfilling) & \cellcolor[HTML]{32CB00}beneficial                     \\ 
                                     & 1 (treat everyone)       & \textless 0 (not self-fulfilling) & \cellcolor[HTML]{FE996B}harmful \\ \hline
                                     & 0 (treat no one)       & \textgreater 0 (self-fulfilling) & \cellcolor[HTML]{32CB00}beneficial                     \\
                                     & 0 (treat no one)       & \textless 0 (not self-fulfilling) & \cellcolor[HTML]{FE996B}harmful \\
\multirow{-2}{*}{desirable (treat low risk patients)} & 1 (treat everyone)       & \textgreater 0 (self-fulfilling) & \cellcolor[HTML]{FE996B}harmful                       \\
                                     & 1 (treat everyone)       & \textless 0 (not self-fulfilling) & \cellcolor[HTML]{32CB00}beneficial 
\end{tabular}
\caption{Overview of when OPM deployment was harmful, based on three pieces of information that are available post-deployment.
This table excludes the trivial case where there is no change in the $Y|X$ distribution post-deployment (see Theorem \ref{thm:calibration_x}).
        $\pi_0$: historical treatment policy (either treat everyone or treat no one in our setup); $AUC(\pi_f)$: AUC in distribution post deployment; $AUC(\pi_0)$: AUC in distribution pre deployment; OPM: outcome prediction model}
\label{tab:results}
\end{table}

In examining such results, one may wonder whether these properties occur in realistic scenarios or only in extreme circumstances. To answer this question, we conducted a numerical experiment via the following data distributions

\begin{align}
    x &\sim B(p_x) \\
    t &\in \{0,1\}\\
    \eta &= \beta_0 + \beta_x x + \beta_t t + \beta_{xt} x t \\
    y &\sim B(\sigma(\eta))
\end{align}

where $p_x$ is the proportion of data points with a positive attribute $X$, $t$ is the historical treatment policy (which is always 0 or 1 according to our assumptions), $B$ the bernoulli distribution, $\sigma$ the sigmoid function and the $\beta$ parameters encode the effects of different components on the outcome. Setting  these parameters, enforcing the assumptions of the theorem and deciding whether a higher $Y$ is better (e.g. in survival) or worse (e.g. for a risk), gives enough information to describe the pre- and post-deployment scenarios. Note that a non-constant policy $\pi_f$ entails that different treatments are now prescribed for the two groups defined by $X$, thus further assumptions on the model and threshold $\lambda$ are not needed. This allows us to calculate discrimination statistics pre- and post-deployment and to determine whether the new policy is harmful. By repeating the experiment for several values of the parameters -- within reasonable ranges -- one can investigate when harmful self-fulfilling policies arise.

\begin{figure}[ht!]
    \centering
    \includegraphics[width=\textwidth]{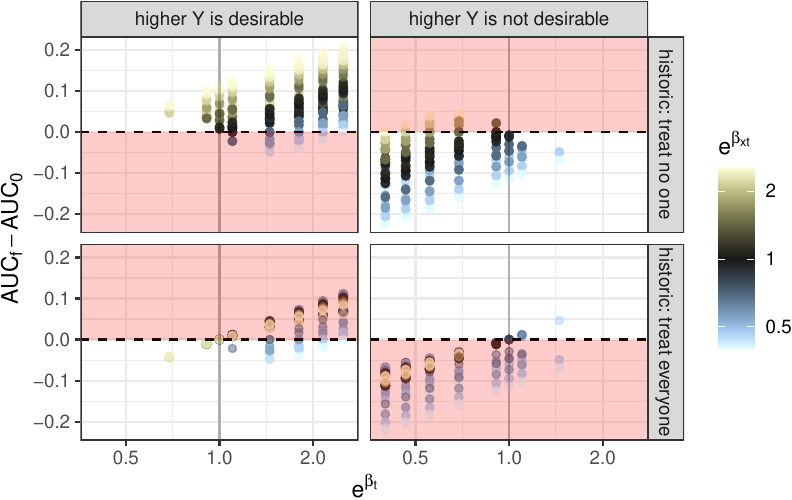}
    \caption{Results of the numerical experiment. Treatment effect is reported on the x-axis on the odds-ratio scale, while the difference in AUC pre- and post-deployment is given on the y-axis. When said difference is positive, we have a self-fulfilling policy. The four panels reflect the different combinations of historical policy and outcome interpretation.  The areas marked  in red denote a harmful effect. Points are color-coded with the value of treatment effect interaction, again with an odds-ratio.}
    \label{fig:simulation_treatment_effect}
\end{figure}

The results match the theoretical findings, and furthermore display that harmful self-fulfilling policies do occur in `common' circumstances. Figure \ref{fig:simulation_treatment_effect} for example, shows several instances of the experiment. A positive difference in AUC denotes a self-fulfilling policy, while harmful policies fall within a red area. Inspection of the figure reveals several scenarios to be harmful and self-fulfilling in the top-right and bottom-left panels. These scenarios can occur at different values of treatment effect (parameter $\beta_t$), and can even lead to an increase of AUC of $>0.1$. For Figure \ref{fig:simulation_treatment_effect} we kept only settings where the treatment effect is beneficial on average. This removes several cases of harmful-selffulfilling prophecies, but is more realistic as treatments are generally only allowed on the market if their average effectiveness is demonstrated in RCTs. In Figure \ref{fig-bt-vs-diff-all} in Appendix \ref{app:numerical_exp} all settings are presented. Furthermore, Figure \ref{fig-bxt-vs-diff} in Appendix \ref{app:numerical_exp} gives another visualization of the same experimental results, this time highlighting that harmful self-fulfilling prophecies occur in the absence of strong treatment effect interactions (i.e. treatment effect heterogeneity, the parameter $\beta_{xt}$). 
Full details on the setup of the numerical experiment and further results can be found in Appendix \ref{app:numerical_exp} and code to reproduce the results is available in the supplemental materials.

Note that the table and the figures highlight a dual problem to `harmful self-fulfilling', which we could call `beneficial self-defeating': the case where AUC decreases post-deployment but the new policy is in fact beneficial. In this case, an over-reliance on performance metrics might lead to another ill-advised decision: the withdrawal of a policy that was in fact beneficial.

\subsection{OPMs that are calibrated pre- and post-deployment are not useful for treatment decisions}
Monitoring discrimination post-deployment and naively interpreting good post-deployment discrimination as a safe deployment is thus not a good strategy, as self-fulfilling prophecies have good post-deployment discrimination but can still be harmful depending on the context. Conversely, beneficial policies may have decreased post-deployment discrimination due to the desirable effect of improving patient outcomes.
We now turn to another key metric of OPMs predicting the risk of an outcome: \emph{calibration} \citep{alba2017discrimination, huang2020tutorial,van2019calibration} and investigate how post-deployment calibration relates to harmful policies.
We use the following definition of calibration.

\begin{definition}\label{def:calibration}
	Let $p(X,Y)$ be a joint distribution over feature $X$ and binary outcome $Y$, and $f: \mathcal{X} \to [0,1]$ an OPM. \emph{$f$ is calibrated with respect to $p(X,Y)$} if, for all  $\alpha \in [0,1]$ in the range of $f$, $\EX_{X,Y \sim p(X,Y)}[Y|f(X)=\alpha]=\alpha$.
\end{definition}

We distinguish two distributions $p_i(Y=1|X)$ on which an \gls{opm} can be calibrated depending on the treatment policy indicated with $i \in \{0,f\}$.
Theorem \ref{thm:main} states that harmful \glspl{opm} can have good pre- and post-deployment discrimination, but can they also have good calibration?
The following theorem shows that OPMs that are calibrated pre- and post-deployment do not lead to better treatment decisions.

\begin{theorem}\label{thm:calibration_x}
Let $f$ be an OPM that is calibrated on historical data and $\pi_f$ be non constant. Such \gls{opm} is calibrated on the deployment distribution iff for every $x \in \mathcal{X}$:
\begin{equation}
    \pi_0(x) = \pi_f(x) \text{ \emph{or} } p(Y_1=1|X=x) = p(Y_0=1|X=x)
\end{equation}
\end{theorem}

Note that this entails that for all $x \in \mathcal{X}$ either the treatment policy does not change, or it changes where it is irrelevant because for that value of $X$ the treatment effect is zero.
Both cases imply the implementation of the \gls{opm} is inconsequential. This may seem  counterintuitive, but an OPM being calibrated both before and after deployment means the distribution has not changed, so the policy remains the same or the policy was changed where it is irrelevant (i.e. no treatment effect). So an OPM that is calibrated on the development cohort, which remains calibrated post deployment is not a useful OPM.

\section{Related work}
\label{sec:relatedwork}

The intuition that deploying models for decision support is an intervention that requires causal evaluation methods goes back at least to the 90s \citep{cooper1997evaluation}, and previous work noted that prediction accuracy does not equal value for treatment decision making \citep{vickers2006decision, moons_transparent_2015,vanamsterdamAlgorithmsActionImproving2024}. Here we take the additional step of exactly characterizing the set of prediction models that yield harmful self-fulfilling prophecies.
The idea that model deployment changes the distribution and affects model performance was noted in several lines of previous work.
Several authors noted that model performance may degrade over time due to the effect of deployment of the model \citep{lenert_prognostic_2019,sperrin_explicit_2019}, but we study the case where model performance does \emph{not} degrade but the implementation of it still caused harm. And also, degraded discrimination may indicate benefit of the deployment.
\citet{perdomo_performative_2021} and \citet{liley_model_2021} study the setting of performing successive model updates, each time after deploying the previous model for decision making.
\citet{perdomo_performative_2021} study when over successive deployments predictive performance stabilizes or reaches optimality, and \citet{liley_model_2021} study both model stability and the effect of model deployment on outcomes.
Our work may be seen as a special case of these works with only a single model deployment and no model update, but we add new insights as we describe exactly \emph{when} a single model deployment leads to harm and good post-deployment discrimination.
Several groups have studied out-of-distribution generalization and its connections to causality and invariance \citep{arjovsky_invariant_2020,wald_calibration_2021,puli_out--distribution_2023} with the aim of removing a model's dependency on \emph{spurious correlations}.
Again our work differs as we are interested in characterizing model performance following a very specific distribution change (a treatment policy change induced by a prediction model) and our main concern is the effect of this policy change on outcomes.
Finally, current guidelines on prediction model validation and deployment focus on discrimination and calibration only, not on these newer invariance metrics \citep{moons_transparent_2015,kattan_american_2016}.
Concurrent work studies the same setup as ours through the lens of domain adaptation, where each (pre) deployment setting is formalized as a domain \cite{boekenEvaluatingCorrectingPerformative2024}.
They describe when the effect of deploying (or updating) an OPM for decision support can be estimated without observing outcomes under the target domain, however both the assumptions and the results diverge from the present work.

We are not the first to warn against naively using OPMs for decision support (see e.g. points 6.3 and 6.7 in \cite{asselGuidelinesReportingStatistics2019}). However, (intended) misuse of OPMs is still far too common in medical research and guidelines, and the reason why this can lead to harmful situations is not well-understood. Our work provides a formal framework to understand the risks of using OPMs without proper validation. %

\section{Discussion}
\label{sec:discussion}
We showed how \glspl{opm} can be harmful self-fulfilling prophecies, meaning they lead to patient harm when used for treatment decision making, but retain good discrimination after deployment. Moreover, we showed that when a model is well calibrated before and after deployment it is not useful for treatment decision making. The upshot of these findings is not only that harmful and self-fulfilling policies exist, but also that in some scenarios it is even \textit{desirable} to see worse discrimination after deployment, since this may signal a beneficial new policy in terms of patient outcomes. These results cast doubt on the adequacy of current practice for the evaluation of predictive models post deployment, when these models are used for decision making.

When interpreting the performance of an OPM post-deployment, a ``high AUC is good, low AUC is bad'' mindset proves to be too simplistic.  A higher performance post-deployment does not necessarily indicate a beneficial policy change, and a lower performance post-deployment is not by itself a sign that the model is harmful. For instance,  the latter may due to poor generalization performance, but also due to the OPM implementation being beneficial and changing the population so that the prediction tasks becomes harder (whence lower AUC), shedding a new light on results such as \citet{wong2021external}. In this second circumstance, removing an OPM-based policy due to low performance would in fact be detrimental in terms of patient outcomes. Our Table \ref{tab:results} can provide concrete guidance for determining whether the new policy was harmful or beneficial. In short, the pre-existing treatment policy, the interpretation of the outcome variable and the change in AUC post-deployment can already give an indication of the effect of the new policy on patient outcomes, provided the assumptions of our settings hold.

In recent years, the United States Food and Drug Administration (FDA) and the European Medical Agency (EMA) have been developing protocols on regulating artificial intelligence based software for medical applications.
The FDA's guiding principles explicitly include a total product life-cycle approach, where post-deployment monitoring and certain potential model updates are foreseen and described during initial approval, both with the aim to ensure post-deployment safety for example under dataset shifts, but also to avoid the need for re-approval after each model update.
Though their guiding principles on `good machine learning practice' \cite{fdaGoodMachineLearning2021} and `Predetermined Change Control Plans' \cite{fdaPredeterminedChangeControl2023} both mention post-deployment monitoring for safety, the intended monitoring seems to center mostly around predictive performance, which our results demonstrate to be insufficient to protect against harmful self-fulfilling prophecies.
The EMA's `Reflection paper on the use of artificial intelligence in the lifecycle of medicines' also recommends pre-planned monitoring but only of predictive performance \cite{emaReflectionPaperUse}.

Requiring explicit monitoring of changes in patient outcomes over time and changes in treatment policy may in some cases be warranted.
Though monitoring patient outcomes in important pre-determined patient subgroups before and after deployment may detect harmful model deployments, before-after comparisons are plagued by well known biases such as potential concurrent changes in other policies or general time-trends in outcomes.
The best experiment to demonstrate the safety of deploying an \gls{opm} is to conduct a cluster randomized controlled trial, where some care-givers are randomly selected to have access to the \gls{opm} and others are not.
The difference in average outcomes of patients between the care-givers with and without access determines whether using the \gls{opm} led to better patient outcomes. When cluster randomized trials are unfeasible, other smaller clinical studies might be the next best option \citep{vasey2022reporting}.
Beyond deployment, how to pre-specify safe model monitoring and updates in a total product life-cycle approach in light of our self-fulfilling prophecy framework is left for future work.

Finally, we note that developing OPMs that ignore the historic treatment policy is in many cases not the optimal approach when the ultimate aim is to improve the policy for assigning treatments.
Instead, researchers should consider using methods developed for improving decisions such as prediction-under-intervention models or models of the conditional average treatment effect (CATE).

Some limitations remain, encoded in the assumptions of our formal results. The setting we describe is kept simple on purpose, a choice that helps in pinpointing the problem but limits somewhat the applicability of this theory to real-world use cases.
The extension of our results to other feature types (continuous and categorical $X$), non-threshold based policies, or to a $\pi_0$
that is not constant (i.e. varies with $X$) or is non-deterministic, is left to future work. Other more complex use cases worth investigating might display policies that are harmful for subgroups identified by variables not included in the list of predictors of the model.
The continuation of this line of work entails the re-evaluation of the metrics to monitor and assess a model's effectiveness, and given that model deployments for decision support are interventions, this will benefit from using the language of causal inference.

\section{Conclusion}
Outcome prediction models can yield harmful self-fulfilling prophecies when used for decision making. %
The current paradigm on prediction model development, deployment and monitoring needs to shift its primary focus away from predictive performance and instead towards changes in treatment policy and patient outcomes.

\printbibliography

\pagebreak

\section{Statements and Declarations}

\subsection{Funding}

The authors declare that no funds, grants, or other support were received for the preparation of this manuscript.

\subsection{Competing Interests}

The authors have no relevant financial or non-financial interests to disclose.

\subsection{Author Contributions}

Study conception and design: WA and GC Manuscript drafting: WA and GC Manuscript revision and editing: all authors

\subsection{Ethics approval}

No ethics approval was sought as no patient data was involved in this study.

\subsection{Consent to participate}

Not applicable

\subsection{Consent to publish}

Not applicable

\appendix

\section{Hypothetical example of a harmful self-fulfilling prophecy}
\label{app:example}

We now give a full-fledged hypothetical example based on realistic assumptions that would result in an OPM yielding a policy that is both harmful and self-fulfilling.

Consider the problem of selecting a subset of end-stage cancer patients for palliative radiotherapy.
Such treatment has severe side-effects and thus domain experts advise to attempt to reduce over-treatment in the population of cancer patients.
To comply with this advice, a medical center needs to decide which patients will not be eligible anymore for the therapy.

The medical center decides to give the therapy to patients with the longest expected overall survival, under the assumption that these patients would be those for whom the side-effects are justifiable.
To support this policy, researchers built an \gls{opm} to predict the probability of 6-months overall survival based on pre-treatment tumor growth rate using historical patient records from the medical center.
Fast-growing tumors are more aggressive so these patients have a shorter survival overall.
The medical center decides to use this model to allocate the therapy and tests the model's discrimination post deployment.
Based on this we have the following facts:

\begin{enumerate}
    \item $X=1$: fast growing tumor, $X=0$: slow-growing tumor;
    \item $\pi_0(X) = 1$, the historical policy was treating everyone;
    \item $p(Y_0=1|X=0) - p(Y_0=1|X=1) > 0$, with radiotherapy, patients with fast growing tumors live shorter
\end{enumerate}

A model with a good fit to the data will predict that patients with slow-growing tumors have a higher probability of 6-months survival. We also assume that the new policy is non-constant and favors those with highest predicted outcome, which means that the new policy will be `treat patients with slow growing tumors but not those with fast growing tumors':
\begin{equation*}
    \pi_f(X) = 1 - X
\end{equation*}

However, it is well known that fast-growing tumors respond better to radiotherapy than slow growing tumors \citep{breurGrowthRateRadiosensitivity1966}. 
Based on this we add the following two assumptions:

\begin{enumerate}
    \item $p(Y_0=1|X=0) - p(Y_1=1|X=0) = 0$, radiotherapy is not effective against slow growing tumors;
    \item $\delta:= p(Y_0=1|X=1) - p(Y_1=1|X=1) < 0$, radiotherapy \emph{is} effective for fast growing tumors.
\end{enumerate}

This means that the antecedent of Proposition \ref{thm:selffulfilling} is satisfied, meaning that $f$ yields a self-fulfilling prophecy in combination with any threshold $\lambda$ such that the resulting policy is non-constant. Removing the therapy from the group $X=1$ will worsen their outcomes by $\delta$, separating the two groups even more and resulting in higher AUC post-deployment.

Moreover, according to the first case of Proposition \ref{thm:self_harmful}, the \gls{opm} is harmful because the new treatment policy leads to worse outcomes for the group with fast growing tumors ($X=1$).
So the \gls{opm}-based policy treats exactly the wrong patients: those who do not benefit from treatment still receive it, those who would benefit from treatment do not, but paradoxically it has good discrimination before and after deployment.

\section{Proofs of main results}\label{apd:proofs}

\subsection{Proof of Proposition \ref{thm:selffulfilling}.}

\begin{proof}

First we give some elementary definitions and equalities. Define
\begin{align}\label{eq:mui}
	\mu_i(x) &= p_i(Y=1|X=x) = \left(1- \pi_i(x) \right) p(Y_0=1|X=x) + \pi_i(x) p(Y_1=1|X=x)
\end{align}

So by the law of total probability we can write
\begin{align}
	p_i(Y=1) &= p_i(X=0) \mu_i(0) + p_i(X=1) \mu_i(1)
\end{align}
By Bayes rule we have:

\begin{equation}
	\label{eq:pxy}
	p_i(X=x|Y=y) = \frac{p_i(Y=y|X=x) p(X=x)}{p_i(Y=y)}
\end{equation}

Filling in the definition of $\mu_i(x)$ into \ref{eq:pxy} using the assumption that $p_i(X=x) = p(X=x)$ we have in particular:

\begin{align}
	\label{eq:px1}
	p_i(X=x|Y=1) &= \frac{\mu_i(x) p(X=x)}{p_i(Y=1)}
\end{align}

ROC-curves are created by transforming a continuous-valued function to a binary prediction based on a varying \emph{threshold} $\tau$ and calculating the \emph{sensitivity} and \emph{specificity} for each value of $\tau$:

\begin{align}
    \label{eq:sens_spec}
    \text{sensitivity} &= p(f(X) \geq \tau | Y=1)  \\
    \text{specificity} &= p(f(X) < \tau | Y=0) 
\end{align}

For each possible threshold, all predictions under the threshold are labeled \emph{negative} and all predictions greater or equal to the threshold \emph{positive}.
In the case of a binary $X$, $f(X)$ only takes two unique values so the ROC-curve is given by just three points:

\begin{enumerate}
    \item sensitivity = 1, specificity = 0 ($\tau = - \infty$)
    \item sensitivity = 0, specificity = 1 ($\tau = + \infty$)
    \item sensitivity = sens, specificity = spec ($\tau = \max_X f(X)$)
\end{enumerate}

See Figure \ref{fig:auc}.
We can directly calculate the AUC by dividing the area under the ROC-curve in two adjacent non-overlapping triangles.
This gives us the following expression for the AUC (see also \cite{muschelliROCAUCBinary2020}):

\begin{figure}
    \centering
    \includegraphics[width=0.5\textwidth]{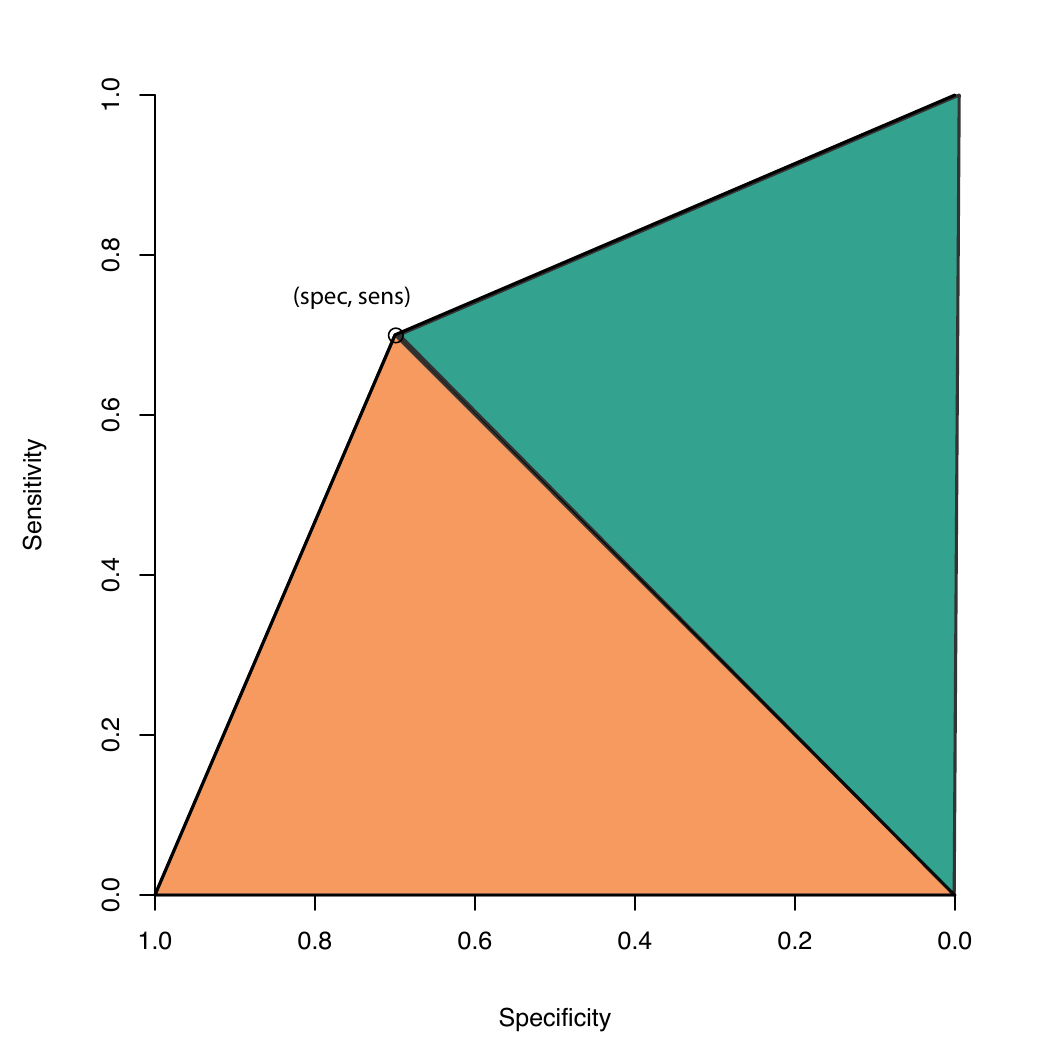}
    \caption{AUC for a binary predictor $X$}
    \label{fig:auc}
\end{figure}

\begin{equation}
    \label{eq:auc}
    \text{AUC} = \textcolor{Orange}{\frac{1}{2} \text{sens}} + \textcolor{OliveGreen}{\frac{1}{2} \text{spec}}
\end{equation}

In this binary case, the area-under the ROC curve is thus determined by a single point denoted as (spec,sens).
A pair $(f,\lambda)$ is self-fulfilling when:

\begin{equation}
    \label{eq:auc_diff}
    \text{AUC}(f) - \text{AUC}(0) = \frac{1}{2} \left( \text{sens}_f + \text{spec}_f - \text{sens}_0 - \text{spec}_0 \right) \geq 0
\end{equation}

We structure the proof by first creating an enumeration over all possible scenarios.
We assumed $\pi_f$ is non-constant, which implies that $f$ varies with $X$.
Since $X$ is binary, it must be that either $f(0)>f(1)$ or $f(1)>f(0)$.
These cases are symmetric under relabeling of $X$ so without loss of generality we proceed assuming that $f(0)>f(1)$ is the case.
Since $\pi_f$ is not constant but $\pi_0$ is, it must be that either the treatment policy changes for $X=0$ but remains the same for $X=1$, or vice versa.
This in turn implies that either $\mu_f(0) = \mu_0(0)$ or $\mu_f(1) = \mu_0(1)$.

To provide a proof for the theorem, we enumerate all the subcases based on two factors:
\begin{enumerate}
	\item for which group does the policy change ($X=0$ or $X=1$)?
	\item for the group with the policy change, does the outcome under the new policy remain the same (the policy is inconsequential as the treatment effect is zero), increase or decrease (this will be beneficial or detrimental depending on whether $Y=1$ is good or bad )
\end{enumerate}

This leads to the following 6 cases:

\begin{itemize}
	\item policy change for which $X$?
		\begin{enumerate}
  \setcounter{enumi}{-1}
			\item $\pi_f(0) \neq \pi_0(0)$ \\
            \textbf{effect of policy change:}
			\begin{itemize}
				\item [$=$:] $\mu_f(0) = \mu_0(0)$, $\mu_f(1) = \mu_0(1)$
				\item [$<$:] $\mu_f(0) < \mu_0(0)$, $\mu_f(1) = \mu_0(1)$
				\item [$>$:] $\mu_f(0) > \mu_0(0)$, $\mu_f(1) = \mu_0(1)$
			\end{itemize}
			\item $\pi_f(1) \neq \pi_0(1)$ \\
            \textbf{effect of policy change:}
			\begin{itemize}
				\item [$=$:] $\mu_f(0) = \mu_0(0)$, $\mu_f(1) = \mu_0(1)$
				\item [$<$:] $\mu_f(0) = \mu_0(0)$, $\mu_f(1) < \mu_0(1)$
				\item [$>$:] $\mu_f(0) = \mu_0(0)$, $\mu_f(1) > \mu_0(1)$
			\end{itemize}
		\end{enumerate}
\end{itemize}

These 6 combinations cover all possibilities.
Since we have that $f(0)>f(1)$, by assumption of a non-deterministic $\pi_f(x) = I_{f(x) > \lambda}$ it must be that for all subcases $\pi_f(0) = 1$ and $\pi_f(1)=0$.
Each of these cases have implications for $\pi_0$ and, depending on which policy changes, $p(Y_1=1|X=0)-p(Y_0=1|X=0)$ or $p(Y_1=1|X=1)-p(Y_0=1|X=1)$.
For instance case $(0,>)$ specifies that $\pi_f(0) \neq \pi_0(0)$ so it follows that $\pi_0 = 0$.
And because $Y_1(0) = \mu_f(0)  > \mu_0(0) =Y_0(0)$ it must be that $p(Y_1=1|X=0) - p(Y_0=1|X=0) > 0$, meaning that the treatment increases the outcome for the group with $X=0$.

In the two cases where the outcomes do not change ($(0,=)$ and $(1,=)$), $(f,\lambda)$ is trivially self-fulfilling as nothing changes in the distribution of $X,Y$ so the sensitivity and specificity remain the same.

We first prove self-fulfillingness in cases $(0,>)$ and $(0,<)$:

\paragraph{Case $(0,>)$ and $(0,<)$}

We first address case $(0,>)$, which gives us this information:

\begin{itemize}
	\item $\pi_f(0) \neq \pi_0(0)$
	\item $\mu_f(0) > \mu_0(0)$
	\item $\mu_f(1) = \mu_0(1)$
\end{itemize}

Since $f(0) > f(1)$ we get these sensitivity and specificity:

\begin{align}
    \label{eq:sens_spec_B}
    \text{sens}_i = p_i(f(X) \geq \max(f) | Y=1) &= p_i(X=0 | Y=1) \\
    \text{spec}_i = p_i(f(X) < \max(f) | Y=0) &= p_i(X=1 | Y=0)
\end{align}

with $i\in \{0, f\}$. Plugging this into \ref{eq:auc_diff} yields:

\begin{align*}
    \text{AUC}(f) - \text{AUC}(0) =  \frac{1}{2} \bigl( &p_f(X=0|Y=1) - p_0(X=0|Y=1)\\
    + &p_f(X=1|Y=0) - p_0(X=0|Y=0) \bigr) \\
    = \frac{1}{2} \bigl( &\mu_f(0) \frac{p(X=0)}{p_f(Y=1)} - \mu_0(0) \frac{p(X=0)}{p_0(Y=1)} \\
    + &(1-\mu_f(1)) \frac{p(X=1)}{p_f(Y=0)} - (1-\mu_0(1)) \frac{p(X=1)}{p_0(Y=0)} \bigr)
\end{align*}

where the first equality is by substitution and rearrangement, and the second by Bayes rule. We can determine the sign of this difference based on the sign of two terms:

\begin{align}
    \label{eq:diff_case}
    = \frac{1}{2} \bigl( p(X=0) \bigl(& \textcolor{Orange}{ \frac{\mu_f(0)}{p_f(Y=1)} - \frac{\mu_0(0)}{p_0(Y=1)}} \bigr) \\
    + p(X=1) & \bigl( \textcolor{OliveGreen}{ \frac{1-\mu_f(1)}{p_f(Y=0)} -  \frac{1-\mu_0(1)}{p_0(Y=0)} } \bigr) \bigr)
\end{align}

We write the difference between pre- and post-deployment expected outcome for the group $X=0$ as

\begin{equation}
    \delta := \mu_f(0) - \mu_0(0)
\end{equation}

This gives us

\begin{align}
    p_f(Y=1) &= p(X=1) \mu_f(1) + p(X=0) \mu_f(0) \\
             &= p(X=1) \mu_0(1) + p(X=0) (\mu_0(0) + \delta) \\
             &= p_0(Y=1) + p(X=0) \delta
\end{align}

where the first step is the law of total probability, the second by the definition of $\delta$ and the case information $\mu_f(1) = \mu_0(1)$, and finally again using the law of total probability. Furthermore 

\begin{align}
    p_f(Y=0) &= 1 - p_f(Y=1)  \\
             &= 1 - p_0(Y=1) - p(X=0) \delta  \\
             &= p_0(Y=0) - p(X=0) \delta
\end{align}

where the second step is by our previous calculation and the other two just the property of binary outcomes. We can now determine the signs of the two terms in \ref{eq:diff_case}.

\begin{align}
    \label{eq:term1}
    \text{sign} \bigl[\textcolor{Orange}{ \frac{\mu_f(0)}{p_f(Y=1)} - \frac{\mu_0(0)}{p_0(Y=1)}} \bigr] &= \text{sign} \bigl[\frac{\mu_f(0)p_0(Y=1) - \mu_0(0)p_f(Y=1)}{p_f(Y=1)p_0(Y=1)} \bigr] \\
    &= \text{sign} \bigl[ \mu_f(0)p_0(Y=1) - \mu_0(0)p_f(Y=1) \bigr]
\end{align}

The first equality is cross-multiplying, the second equality is because the product of two probabilities (which are positive by assumption) is always a positive number.

Filling in the definition of $\delta:$

\begin{align}
    \MoveEqLeft \text{sign} \bigl[\textcolor{Orange}{ \frac{\mu_f(0)}{p_f(Y=1)} - \frac{\mu_0(0)}{p_0(Y=1)}} \bigr] \\
    &= \text{sign} \bigl[ (\mu_0(0) + \delta) p_0(Y=1) - \mu_0(0) (p_0(Y=1) + p(X=0) \delta) \bigr] \\
    &= \text{sign} \bigl[ \delta p_0(Y=1) - \mu_0(0) p(X=0) \delta \bigr] \\
    &= \text{sign} \bigl[ \delta (p_0(Y=1) - \mu_0(0) p(X=0)) \bigr] \\
    &= \text{sign} \bigl[ \delta \mu_0(1) p(X=1) \bigr] \\
    &= \text{sign} \bigl[\textcolor{Orange}{\delta} \bigr]
\end{align}

In the second equality we remove canceling terms.
In the third equality we pull out $\delta$.
In the fourth equality we use the expansion of $p_0(Y=1) = p(X=0)\mu_0(0) + p(X=1)\mu_0(1)$,
and for the final equation we note again that $\mu_0(1)$ and $p(X=1)$ are positive probabilities so the sign is determined by the sign of $\delta$.

Now for the second term of \ref{eq:diff_case}:

\begin{align}
    \text{sign} \bigl[\textcolor{OliveGreen}{ \frac{1-\mu_f(1)}{p_f(Y=0)} -  \frac{1-\mu_0(1)}{p_0(Y=0)}} \bigr] &= \text{sign} \bigl[ \frac{1-\mu_0(1)}{p_f(Y=0)} -  \frac{1-\mu_0(1)}{p_0(Y=0)} \bigr] \\
    &= \text{sign} \bigl[\bigl(1-\mu_0(1) \bigr) \bigl(\frac{1}{p_f(Y=0)} - \frac{1}{p_0(Y=0)} \bigr) \bigr] \\
    &= \text{sign} \bigl[\frac{1}{p_f(Y=0)} - \frac{1}{p_0(Y=0)}  \bigr] \\
    &= \text{sign} \bigl[\frac{p_0(Y=0) - p_f(Y=0)}{p_f(Y=0)p_0(Y=0)}  \bigr] \\
    &= \text{sign} \bigl[p_0(Y=0) - p_f(Y=0) \bigr] \\
    &= \text{sign} \bigl[p_0(Y=0) - p_0(Y=0) + p(X=0) \delta \bigr] \\
    &= \text{sign} \bigl[p(X=0) \delta \bigr] \\
    &= \text{sign} \bigl[\textcolor{OliveGreen}{\delta} \bigr]
\end{align}

The first equality uses the case assumption that $\mu_f(1) = \mu_0(1)$.
The second equality pulls out the common term $(1-\mu_0(1))$.
The third equality follows because $0 < \mu_0(1) < 1$.
The fourth and fifth equality are cross-multiplying and again using the positive probability property.
In the sixth equality we substitute in the definition of $\delta$.
The seventh equality removes the canceling terms, and the final equality again relies on that $0<p(X=0)$.

So both terms in \ref{eq:diff_case} have the sign of $\delta$.
In subcase $(0,>)$ $\delta$ has positive sign, so
\begin{equation*}
    \text{AUC}(f) - \text{AUC}(0) > 0
\end{equation*}

and $(f,\lambda)$ is self-fulfilling.

Immediately it is clear that in subcase $(0,<)$, $(f,\lambda)$ is not self-fulfilling, as subcase $(0,<)$ equals subcase $(0,>)$ in all respects except that instead it has a negative sign for $\delta$.

\paragraph{Case $(1,>)$ and $(1,<)$}

We first address case $(1,>)$, which gives us this information:

\begin{itemize}
	\item $\pi_f(1) \neq \pi_0(1)$
	\item $\mu_f(0) = \mu_0(0)$
	\item $\mu_f(1) > \mu_0(1)$
\end{itemize}

Again we write the difference between pre- and post-deployment expected outcome as $\delta$, this time for the group $X=1$:

\begin{equation}
    \delta := \mu_f(1) - \mu_0(1)
\end{equation}

This gives us

\begin{align}
    p_f(Y=1) &= p(X=1) \mu_f(1) + p(X=0) \mu_f(0) \\
             &= p(X=1) (\mu_0(1) + \delta) + p(X=0) \mu_0(0) \\
             &= p_0(Y=1) + p(X=1) \delta
\end{align}

where the first step is the law of total probability, the second by the definition of $\delta$ and the case information $\mu_f(0) = \mu_0(0)$, and finally again using the law of total probability. Furthermore 

\begin{align}
    p_f(Y=0) &= 1 - p_f(Y=1)  \\
             &= 1 - p_0(Y=1) - p(X=1) \delta  \\
             &= p_0(Y=0) - p(X=1) \delta
\end{align}

where the second step is by our previous calculation and the other two just the property of binary outcomes. We can now determine the signs of the two terms in \ref{eq:diff_case}.

The first two steps for the first are the same as in the case $(0,>)$ (see Equation \ref{eq:term1}), after these steps we substitute in the new definition of $\delta$:

\begin{align}
    \MoveEqLeft \text{sign} \bigl[\textcolor{Orange}{ \frac{\mu_f(0)}{p_f(Y=1)} - \frac{\mu_0(0)}{p_0(Y=1)}} \bigr] \\
    &= \text{sign} \bigl[ \mu_f(0)p_0(Y=1) - \mu_0(0)p_f(Y=1) \bigr] \\
    &= \text{sign} \bigl[ \mu_0(0) p_0(Y=1) - \mu_0(0) (p_0(Y=1) + p(X=1) \delta) \bigr] \\
    &= \text{sign} \bigl[ - \mu_0(0) p(X=0) \delta \bigr] \\
    &= \text{sign} \bigl[\textcolor{Orange}{ - \delta} \bigr]
\end{align}

In the third equality we remove canceling terms.
For the final equation we note again that $\mu_0(0)$ and $p(X=0)$ are positive probabilities so the sign is determined by the sign of $\delta$.

Now for the second term of \ref{eq:diff_case}:

\begin{align}
    \MoveEqLeft \text{sign} \bigl[\textcolor{OliveGreen}{ \frac{1-\mu_f(1)}{p_f(Y=0)} -  \frac{1-\mu_0(1)}{p_0(Y=0)}} \bigr]\\
    &= \text{sign} \bigl[ \frac{(1-\mu_f(1))p_0(Y=0) - (1-\mu_0(1))p_f(Y=0)}{p_f(Y=0)p_0(Y=0)} \bigr] \\
    &= \text{sign} \bigl[ (1-\mu_f(1))p_0(Y=0) - (1-\mu_0(1))p_f(Y=0) \bigr] \\
    &= \text{sign} \bigl[ (1-(\mu_0(1)+ \delta))p_0(Y=0) - (1-\mu_0(1))(p_0(Y=0)-p(X=1)\delta) \bigr] \\
    &= \text{sign} \bigl[ -\delta p_0(Y=0) - (1-\mu_0(1))(-p(X=1)\delta) \bigr] \\
    &= \text{sign} \bigl[ -\delta (p_0(Y=0) - (1-\mu_0(1))p(X=1)) \bigr] \\
    &= \text{sign} \bigl[ -\delta ((1-\mu_0(0))p(X=0)) \bigr] \\
    &= \text{sign} \bigl[\textcolor{OliveGreen}{-\delta} \bigr]
\end{align}

The first equality uses cross-multiplication to gather the sum.
The second equality follows because we're dividing by a positive number.
The third equality is filling in the definition on $\delta$.
The fourth equality removes canceling terms.
The fifth equality factors out $-\delta$.
The seventh equality is by the law of total probability.

So both terms in \ref{eq:diff_case} have the sign of $-\delta$.
In subcase $(1,>)$ $\delta$ has positive sign, so
\begin{equation*}
    \text{AUC}(f) - \text{AUC}(0) < 0
\end{equation*}

and $(f,\lambda)$ is not self-fulfilling.

Immediately it is clear that in subcase $(1,<)$, $(f,\lambda)$ is self-fulfilling, as subcase $(1,<)$ equals subcase $(1,>)$ in all respects except that instead it has a negative sign for $\delta$.

\paragraph{Enumerating all the cases}

As said, in the two cases where the outcomes do not change ($(0,=),(1,=)$), $(f,\lambda)$ is trivially self-fulfilling.

Putting all the pieces of information for all subcases together in Table \ref{tab:auc_proof} we see that when $p(Y_1=1|X=x) - p(Y_0=1|X=x) \geq 0$ (the treatment effect is never negative), $(f,\lambda)$ is self-fulfilling.
Also, when $p(Y_1=1|X=x)-p(Y_0=1|X=x) < 0$ (the treatment effect is always negative), $(f,\lambda)$ is never self-fulfilling.
These observations conclude the proof.

\begin{table}[]
    \centering
	\begin{tabular}{llllllll}
		\multicolumn{2}{l}{subcase} & $\pi_0$ & $\pi_f(0)$ & $\pi_f(1)$ & CATE(0) & CATE(1) & self-fulfilling  \\ \cline{1-8}
		 0      & $=$      & 0       & 1          & 0          & 0               &                 & yes                \\
		 0      & $<$      & 0       & 1          & 0          & -               &                 & no                \\
		 0      & $>$      & 0       & 1          & 0          & +               &                 & yes                \\
		 1      & $=$      & 1       & 1          & 0          &                 & 0               & yes                \\
		 1      & $<$      & 1       & 1          & 0          &                 & +               & yes                \\
		 1      & $>$      & 1       & 1          & 0          &                 & -               & no                
	\end{tabular}
	\caption{Enumeration of all possible subcases. The first column indicates for which value of $X$ the treatment policy changes. The second column indicates whether this change improves outcomes for that group ($>$), reduces outcomes ($<$) or is irrelevant ($=$).
 $+/-$ indicates the sign of the subgroup treatment effect CATE$(x) := p(Y_1=1|X=x)-p(Y_0=1|X=x)$;
 }
	\label{tab:auc_proof}
\end{table}
	
\end{proof}

\subsection{Proof of Proposition \ref{thm:self_harmful}.}

Given that we assumed binary $T$ and $X$, we can write the expected value of the outcome conditional on these two variables with four parameters without making parametric assumptions, marginalizing over other variables different than $X$ and $T$.
For ease of interpretation of our results we write the expected value as a sum:
\begin{equation}
	\label{eq:dgm}
p(Y_{T=t}=1|X=x) = \alpha + \beta_x x + \beta_t t + \beta_{xt} x t
\end{equation}
Note that this is not an assumption on the generating process of the outcome Y, which could have arbitrary form, it is only a formal device to represent the four outcomes of interest, one for each value of $X$ and $T$.

We now proceed to prove the Proposition for the case where higher outcome is better; to obtain a proof for the symmetric case (higher outcome is worse) one needs only to switch the sign in the inequalities \ref{toshow:harmful} and \ref{toshow:harmful2}, along with their specialization in the subcases.
\begin{proof}
A treatment is harmful for the group with $X=x'$ iff $p_f(Y=1|X=x') < p_0(Y=1|X=x')$, where according to definition \ref{eq:calibration_train} $p_i(Y = 1|X) = \EX_{T \sim \pi_i(X)} p(Y_T=1|X)$ 
The proof continues as a case distinction depending on the value of $x'$.

\paragraph{Case $x'=1$.} For $x'=1$ the definition of harmful translates to \begin{equation}
\label{toshow:harmful}
\left(\pi_f(1) - \pi_0(1)\right)\left(\beta_t + \beta_{xt}\right) < 0
\end{equation}
We consider the possible values of $\pi_f$ and $\pi_0$ in subcases. Note that if $\pi_f(1)=\pi_0(1)$ the above inequality cannot hold since all terms cancel out and the treatment cannot be harmful (because nothing changes for group $X=1$), so we only consider subcases where these two differ.

\paragraph{Subcase 1.} We have $\pi_f(1) = 0, \pi_f(0) = 1$ and $\pi_0(x) = 1$. In this scenario, we were treating everyone and with the new policy we withhold treatment from group $X=1$. In this case statement \ref{toshow:harmful} specializes to $\beta_t + \beta_{xt}> 0$, meaning that treatment was beneficial and removing it will do damage to group $X=1$.

\paragraph{Subcase 2.} We have $\pi_f(1) = 1, \pi_f(0) = 0$ and $\pi_0(x) = 0$. In this scenario, we were treating nobody and with the new policy we introduce treatment for group $X=1$. In this case statement \ref{toshow:harmful} specializes to $\beta_t + \beta_{xt}< 0$, meaning that treatment is harmful and adding it damages group $X=1$.

\paragraph{Case $x'=0$.} For $x'=0$ the definition of harmful translates to     
\begin{equation}
\label{toshow:harmful2}
\left(\pi_f(0) - \pi_0(0)\right) \beta_t < 0
\end{equation}
Again if $\pi_f(0)=\pi_0(0)$ the above inequality cannot hold since all terms cancel out and the treatment cannot be harmful (because nothing changes for group $X=0$), so we only consider subcases where these two differ.

\paragraph{Subcase 1.} We have $\pi_f(1) = 0, \pi_f(0) = 1$ and $\pi_0(x) = 0$. In this scenario, we were treating nobody and with the new policy we introduce treatment from group $X=0$. In this case the statement \ref{toshow:harmful2} specializes to $\beta_t < 0$, which is what we intended to prove. %

\paragraph{Subcase 2.} We have $\pi_f(1) = 1, \pi_f(0) = 0$ and $\pi_0(x) = 1$. In this circumstance statement \ref{toshow:harmful2} specializes to $\beta_t > 0$.
\end{proof}

\subsection{Proof of Theorem \ref{thm:calibration_x}.}
By assumption $f$ perfectly fits the historical data, so:

\begin{align*}
	f(X=x) =p_0(Y=1|X=x) = \EX_{T \sim \pi_0(x)} p(Y_T=1|X=x).
\end{align*}

We now prove that $f$ is calibrated on the deployment distribution generated by $\pi_f$ iff for all $x \in \mathcal{X}$:

\begin{equation}
    \pi_0(x) = \pi_f(x) \text{ \emph{or} } p(Y_1=1|X=x) = p(Y_0=1|X=x)
\end{equation}

\begin{proof}
    As a shorthand define:
    
    \begin{align*}
        \mu_i(x) \coloneqq &p_i(Y=1|X=x) \\
                = &(1-\pi_i(x))p(Y_0=1|X=x) + \pi_i(x)p(Y_1=1|X=x).
    \end{align*}
    
    $f$ perfectly fits the historical data so:
    
    \begin{align}\label{eq:strong_calib_pi0}
    	f(X=x) = \mu_0(x), \forall x \in \mathcal{X}.
    \end{align}

	$f$ is calibrated on the post-deployment distribution when for all $\alpha \in [0,1]$ in the range of $f$, $\EX_{X,Y \sim p_f(X,Y)}[Y|f(X)=\alpha]=\alpha$.
 So if $f$ is calibrated on both the historic distribution and the post-deployment distribution we have that:
 
\begin{align*}
&\EX_{X,Y \sim p_f(X,Y)}[Y|f(X)=\alpha]
\\
=&\EX_{X,Y \sim p_f(X,Y) |f(X)=\alpha}[Y]
\\
=&\EX_{X,Y \sim p_f(X,Y)}[Y 1[f(X)=\alpha]] / \EX_{X\sim p_f(X)}[ 1[f(X)=\alpha]]
\\
=&\EX_{X,Y \sim p_0(X,Y)}[Y 1[f(X)=\alpha]] / \EX_{X\sim p_0(X)}[ 1[f(X)=\alpha]] 
\end{align*}

Where $1[..]$ is used for the indicator function. We first show that this holds iff for every $x \in \mathcal{X}$, $f(x) = \mu_0(x) = \mu_f(x)$. 
Note that in the last two equations above, the denominators are the same as $p_0(X)=p_f(X)$, so also the enumerators must be the same, so:

\begin{align*}
&\EX_{X\sim p_0(X)} \EX_{Y \sim p_0(Y | X) }[Y 1[f(X)=\alpha]] = \EX_{X\sim p_f(X)} \EX_{Y \sim p_f(Y | X) }[Y 1[f(X)=\alpha]]
\\
\iff&
\EX_{X\sim p_0(X)} 1[f(X)=\alpha] \EX_{Y \sim p_0(Y | X) }[Y] = \EX_{X\sim p_f(X)} 1[f(X)=\alpha] \EX_{Y \sim p_f(Y | X) }[Y]
\\
\iff&
\EX_{X\sim p_0(X)} 1[f(X)=\alpha] 
\EX_{Y_0, Y_1 | X}[(1-\pi_0(X))Y_0 + \pi_0(X)Y_1] 
\\
&= \EX_{X\sim p_f(X)} 1[f(X)=\alpha] \EX_{Y_0, Y_1 | X}[(1-\pi_f(X)Y_0 + \pi_f(X)Y_1]
\end{align*}

Since by assumption $p_0(X) = p_f(X) = p(X)$ we have that

\begin{align*}
\iff&
\EX_{X\sim p(X)} 1[f(X)=\alpha] 
\EX_{Y_0, Y_1 | X}[(1-\pi_0(X))Y_0 + \pi_0(X)Y_1] 
\\
&= \EX_{X\sim p(X)} 1[f(X)=\alpha] \EX_{Y_0, Y_1 | X}[(1-\pi_f(X))Y_0 + \pi_f(X)Y_1]
\\
\iff&
\EX_{X,Y_0, Y_1} 1[f(X)=\alpha] ((1-\pi_0(X))Y_0 + \pi_0(X)Y_1)
\\
&= \EX_{X,Y_0, Y_1} 1[f(X)=\alpha] ((1-\pi_f(X))Y_0 + \pi_f(X)Y_1)
\\
\iff&
\EX_{X} 1[\mu_0(X)=\alpha] \mu_0(X)
= \EX_{X} 1[\mu_0(X)=\alpha] \mu_f(X)
\end{align*}

Where in the last line we substituted the definition of $\mu$ and used the assumption that $f(X) = \mu_0(X)$.
Finally we note that by assumption $\pi_f(X)$ is non-constant. As $X$ is binary it must be that $f$ is injective. 
This implies that the expectation in the last line is given by the value of $\mu$ on a single point corresponding with $\alpha$ which proves that $\mu_0(X)=\mu_f(X)$.

Looking at the difference between $\mu_0(X)$ and $\mu_f(X)$ we see that:

	\begin{align*}
	    \MoveEqLeft \mu_f(X) - \mu_0(X) = \\
     & \left((1-\pi_f(X))p(Y_0=1|X) + \pi_f(X) p(Y_1=1|X) \right) - \\
     &\left((1-\pi_0(X))p(Y_0=1|X) + \pi_0(X) p(Y_1=1|X) \right) \\
	    =& \left( \pi_f(X) - \pi_0(X) \right) \left(p(Y_1=1|X) - p(Y_0=1|X) \right) 
	\end{align*}
    Hence the difference $\mu_f(X) - \mu_0(X)$ is zero iff at least one of the last two terms is zero. This means that $f$ is calibrated on the deployment distribution iff for every $x$ either $\pi_f(x) = \pi_0(x)$ or $p(Y_1=1|X)=p(Y_0=1|X)$
	
\end{proof}

\section{Numerical experiment}
\label{app:numerical_exp}

\subsection{Experimental setup}\label{experimental-setup}

We parameterize the joint distribution with a marginal distribution of
\(X\), a conditional of \(T|X\) and \(Y|T,X\), where we note that by
assumption in the historic distribution, the treatment policy is
\emph{independent} of \(X\), and also that the marginal distribution of
\(X\) does not change after model deployment. Let \(B(.)\) denote the
bernoulli distribution and \(\sigma(x) = \frac{1}{1+e^{-x}}\) the
sigmoid (logistic) function.

\begin{align}
    x &\sim B(p_x) \\
    t &= p_0(T) \in \{0,1\}\\
    \eta &= \beta_0 + \beta_x x + \beta_t t + \beta_{xt} x t \\
    y &\sim B(\sigma(\eta))
\end{align}

The parameter grid is:

\begin{longtable}[]{@{}
  >{\raggedright\arraybackslash}p{(\columnwidth - 6\tabcolsep) * \real{0.1077}}
  >{\raggedright\arraybackslash}p{(\columnwidth - 6\tabcolsep) * \real{0.1077}}
  >{\raggedright\arraybackslash}p{(\columnwidth - 6\tabcolsep) * \real{0.2923}}
  >{\raggedright\arraybackslash}p{(\columnwidth - 6\tabcolsep) * \real{0.4923}}@{}}
\toprule\noalign{}
\begin{minipage}[b]{\linewidth}\raggedright
parameter
\end{minipage} & \begin{minipage}[b]{\linewidth}\raggedright
distribution
\end{minipage} & \begin{minipage}[b]{\linewidth}\raggedright
interpretation
\end{minipage} & \begin{minipage}[b]{\linewidth}\raggedright
values
\end{minipage} \\
\midrule\noalign{}
\endhead
\bottomrule\noalign{}
\endlastfoot
\(p(X=1)\) & \(p(X)\) & marginal distribution of \(X\) & \(0.2, 0.5\) \\
\(p_0(T=1)\) & \(p_0(T)\) & historic treatment policy & \(0,1\) \\
\(\beta_0\) & \(p(Y|T,X)\) & intercept on log odds scale & \(-0.5\) \\
\(\beta_x\) & \(p(Y|T,X)\) & log odds ratio for \(X\) &
log\((1.1,1.45,1.8,2.15,2.5)\) \\
\(\beta_t\) & \(p(Y|T,X)\) & log odds ratio for \(T\) & log\((1/2.5,\)
\(1/2.15,\) \(1/1.8,\) \(1/1.45,\) \(1/1.1,\) \(1,\) \(1.1,\) \(1.45,\)
\(1.8,\) \(2.15,\) \(2.5)\) \\
\(\beta_{xt}\) & \(p(Y|T,X)\) & log odds for interaction between \(T\)
and \(X\) & log\((1/2.5,\) \(1/2.15,\) \(1/1.8,\) \(1/1.45,\) \(1/1.1,\)
\(1,\) \(1.1,\) \(1.45,\) \(1.8,\) \(2.15,\) \(2.5)\) \\
higher \(Y\) & & is higher \(Y\) better or worse & better, worse \\
\end{longtable}

For these parameter values we first calculate the joint probability
\(p(X,Y)\) under the historic distribution.

For some parameter values \(p(Y=1|X=0) = p(Y=1|X=1)\), these settings
are removed as they imply that the outcome risk is independent of \(X\)
in the historical setting and would lead to a constant OPM-based
treatment policy. When then also calculate outcomes under the new
treatment policy with the outcome prediction model.

For each setting we calculate discrimination statistics (sensitivity,
specificity, AUC), using:

\begin{align}
    \text{sens} &= p(f(X) > \lambda|Y=1) \\
    \text{spec} &= p(f(X) < \lambda|Y=0) \\
    \text{AUC} &= \frac{1}{2}(\text{sens} + \text{spec})
\end{align}

See the Appendix B for the derivation of the formula for AUC. First we
need to determine whether the OPM-derived policy \(\pi_f(X) = X\) (when
\(E[Y|X=1]>E[Y|X=0]\)) or \(\pi_f(X) = 1-X\).

Then we compare the AUC before and after deployment to see if \(f\) is
self-fulfilling, and the expected value of \(Y\) before and after
deployment to see whether the new policy was harmful. Note that in our
case harm for a subgroup is equivalent to marginal harm because only the
outcomes change for only one value of \(X\) (see Remark 9).

\subsection{Results}\label{results}

\subsubsection{Experiment checks: Proposition 5 and Table
1}\label{experiment-checks-proposition-5-and-table-1}

With these in hand, we can check Proposition 5 that: i) if the treatment
effect is always positive, then \((f,\lambda)\) is self-fulfilling; ii)
if the treatment effect is always negative, then \((f,\lambda)\) is not
self-fulfilling.

\begin{longtable}[]{@{}
  >{\raggedleft\arraybackslash}p{(\columnwidth - 6\tabcolsep) * \real{0.2212}}
  >{\raggedleft\arraybackslash}p{(\columnwidth - 6\tabcolsep) * \real{0.3558}}
  >{\raggedleft\arraybackslash}p{(\columnwidth - 6\tabcolsep) * \real{0.1923}}
  >{\raggedleft\arraybackslash}p{(\columnwidth - 6\tabcolsep) * \real{0.2308}}@{}}
\toprule\noalign{}
\begin{minipage}[b]{\linewidth}\raggedleft
\(\text{sign}(\beta_t)\)
\end{minipage} & \begin{minipage}[b]{\linewidth}\raggedleft
\(\text{sign}(\beta_{t}+ \beta_{xt})\)
\end{minipage} & \begin{minipage}[b]{\linewidth}\raggedleft
self-fulfilling (N)
\end{minipage} & \begin{minipage}[b]{\linewidth}\raggedleft
not self-fulfilling (N)
\end{minipage} \\
\midrule\noalign{}
\endhead
\bottomrule\noalign{}
\endlastfoot
-1 & -1 & 1508 & 0 \\
-1 & 0 & 100 & 100 \\
-1 & 1 & 200 & 200 \\
0 & -1 & 140 & 40 \\
0 & 0 & 0 & 40 \\
0 & 1 & 0 & 200 \\
1 & -1 & 320 & 44 \\
1 & 0 & 0 & 180 \\
1 & 1 & 0 & 1560 \\
\end{longtable}

We can also see that the numerical experiments follow Table 1.

\begin{longtable}[]{@{}lrlr@{}}
\toprule\noalign{}
higher \(Y\) is & \(p_0(T=1)\) & selffulfilling & harmful (\%) \\
\midrule\noalign{}
\endhead
\bottomrule\noalign{}
\endlastfoot
worse & 0 & true & 1 \\
worse & 0 & false & 0 \\
worse & 1 & true & 0 \\
worse & 1 & false & 1 \\
better & 0 & true & 0 \\
better & 0 & false & 1 \\
better & 1 & true & 1 \\
better & 1 & false & 0 \\
\end{longtable}

\subsubsection{Plots}\label{plots}

\begin{figure}

\centering{

\includegraphics{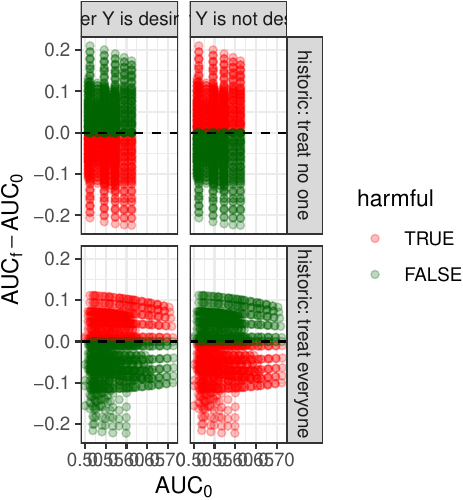}

}

\caption{\label{fig-auc-pre-vs-diff}AUC under historic policy versus AUC
increase under OPM policy, OPM: outcome prediction model}

\end{figure}

\begin{figure}

\centering{

\includegraphics{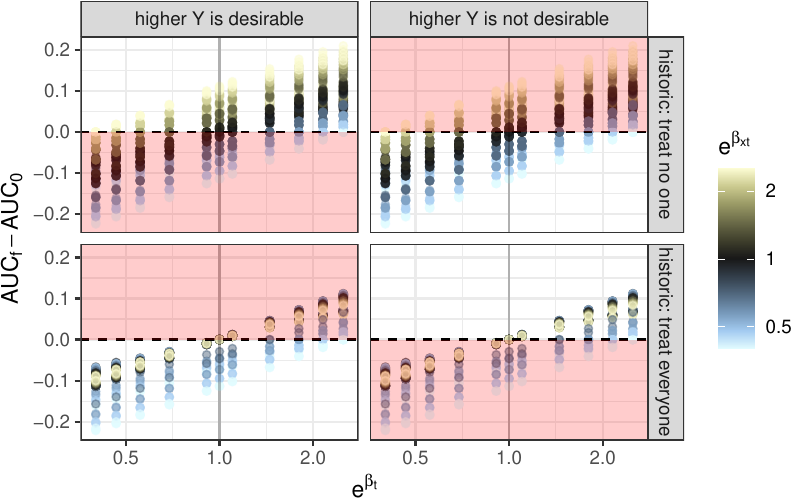}

}

\caption{\label{fig-bt-vs-diff-all}AUC difference versus treatment
effect}

\end{figure}

\begin{figure}

\centering{

\includegraphics{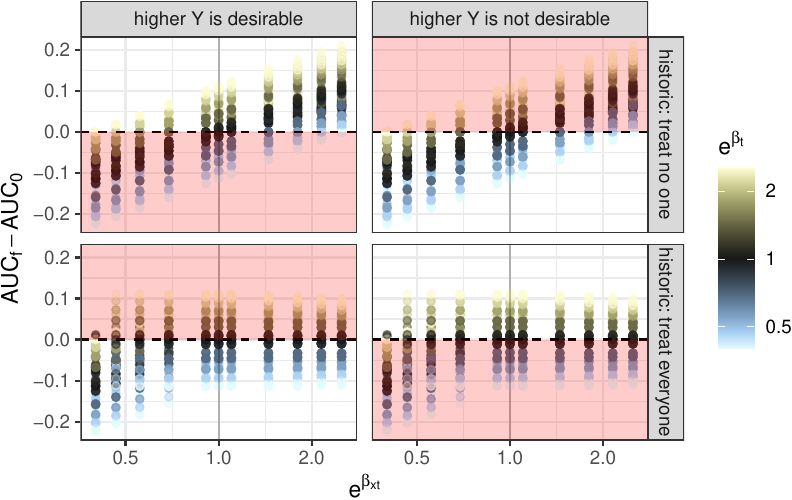}

}

\caption{\label{fig-bxt-vs-diff-all}AUC difference versus treatment
effect interaction}

\end{figure}

We now visualize the results. We include the following plots:

\begin{itemize}
\item
  Figure~\ref{fig-auc-pre-vs-diff}: A scatter plot of AUC on the
  historic data versus AUC in the post-deployment setting
\item
  Figure~\ref{fig-bt-vs-diff-all}: A scatter plot of the odds-ratio for
  treatment \(e^{\beta_t}\) for the group with \(X=0\), versus
  difference in AUC between pre and post deployment, with points colored
  by the treatment effect interaction term \(e^{\beta_{xt}}\), and
  regions indicating whether the OPM-derived policy is harmful or not
\item
  Figure~\ref{fig-bxt-vs-diff-all}: Like Figure~\ref{fig-bt-vs-diff-all}
  but with \(e^{\beta_{xt}}\) on the x-axis and \(e^{\beta_t}\) as color
  codes.
\end{itemize}

\begin{figure}

\centering{

\includegraphics{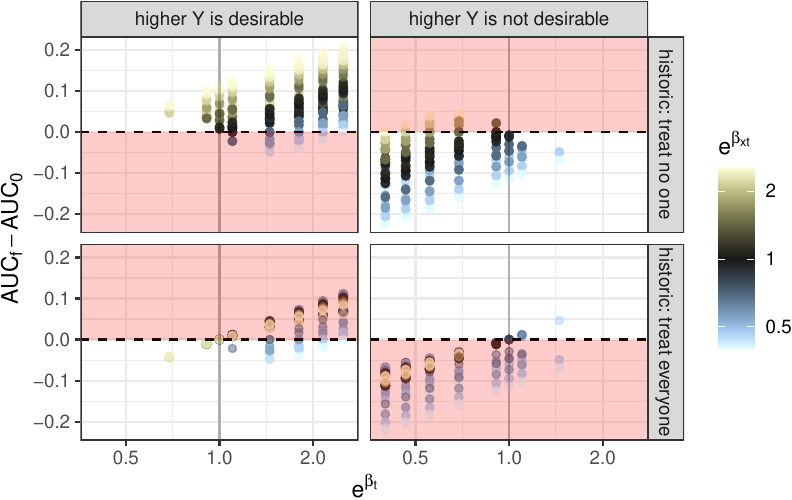}

}

\caption{\label{fig-bt-vs-diff}AUC difference versus treatment effect,
only including settings where treatment is beneficial on average.}

\end{figure}

\begin{figure}

\centering{

\includegraphics{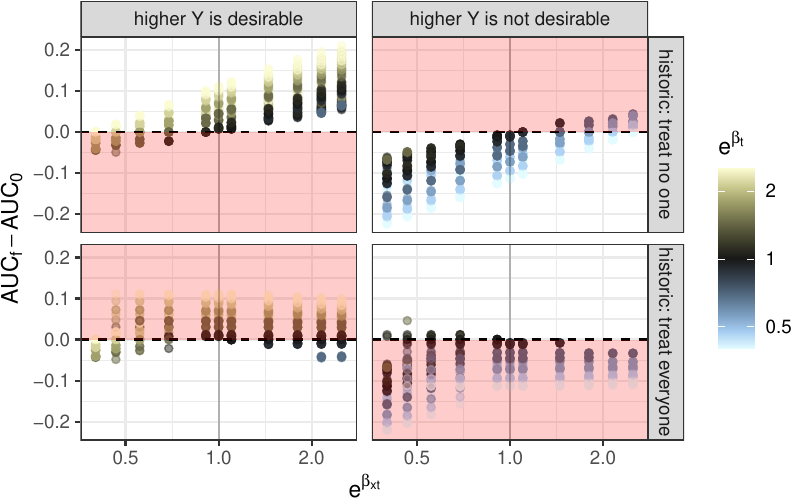}

}

\caption{\label{fig-bxt-vs-diff}AUC difference versus treatment effect
interaction, only including settings where treatment is beneficial on
average.}

\end{figure}

Note that these three Figures include settings where the treatment is
\emph{always detrimental} (e.g.~\(\beta_t >0, \beta_{xt} \ge 0\) and
higher Y is worse). This means that no one should ever be treated with
this treatment, making it highly unlikely that these treatments are in
current clinical use so these settings are not very realistic. Instead,
we subset the settings to those where the treatment is beneficial
\emph{on average}, although it does not have to be effective for both
\(X=0\) and \(X=1\). This is typically the level of evidence available
from RCTs before treatments are allowed on the market. These Figures
are:

\begin{itemize}
\item
  Figure~\ref{fig-bt-vs-diff}: Like Figure~\ref{fig-bt-vs-diff-all} but
  subsetted to settings where treatment is are beneficial on average.
\item
  Figure~\ref{fig-bxt-vs-diff}: Like Figure~\ref{fig-bxt-vs-diff-all}
  but subsetted to settings where treatment is are beneficial on
  average.
\end{itemize}

Figure~\ref{fig-bt-vs-diff} is also presented in the main text.

\end{document}